\documentclass[MSNbibl,number,citesort,dvips]{arxstspdf}
\usepackage{flushend}
\usepackage{stfloats}
\volume{27}
\issue{4}
\pubyear{2012}
\firstpage{447}
\lastpage{449}
\doi{10.1214/12-STS409} 

\makeatletter

\newcommand{\R}{{\mathbb{R}}}
\newcommand{\bbeta}{{\bar{\beta}}}
\newcommand{\hbeta}{{\hat{\beta}}}
\newcommand{\supp}{\operatorname{supp}}

\makeatother

\begin{document}
\begin{frontmatter}

\title{Introduction to the Special Issue on Sparsity and Regularization
Methods}
\runtitle{Introduction}

\begin{aug}
\author[a]{\fnms{Jon} \snm{Wellner}\corref{}\ead[label=e1]{jaw@stat.washington.edu}}
\and
\author[b]{\fnms{Tong} \snm{Zhang}\ead[label=e2]{tzhang@stat.rutgers.edu}}
\runauthor{J. Wellner and T. Zhang}

\affiliation{University of Washington and Rutgers University}

\address[a]{Jon Wellner is Professor of Statistics and Biostatistics, Department of Statistics,
University of Washington, Seattle, Washington 98112-2801, USA \printead{e1}.}
\address[b]{Tong Zhang is Professor of Statistics, Department of Statistics,
Rutgers University, New Jersey, USA \printead{e2}.}

\end{aug}


%

\end{frontmatter}

\section{Introduction}
Traditional statistical inference considers relatively small data sets
and the corresponding theoretical analysis focuses on the asymptotic
behavior of a statistical estimator when the number of samples
approaches infinity.
However, many data sets encountered in modern applications have
dimensionality significantly
larger than the number of training data available, and for such
problems the classical statistical tools become inadequate.
In order to analyze high-dimensional data, new statistical methodology
and the corresponding theory have to be developed.

In the past decade, sparse modeling and the corresponding use of sparse
regularization methods
have emerged as a major technique to handle high-dimen\-sional data.
While the data dimensionality is high, the basic assumption in this approach
is that the actual estimator is sparse in the sense that only a small
number of components are nonzero.
On the practical side, the sparsity phenomenon has been ubiquitously
observed in applications, including signal recovery, genomics,
computer vision, etc. On the theoretical side,
this assumption makes it possible to overcome the problem associated
with estimating more parameters
than the number of observations which is impossible to deal with in the
classical setting.\looseness=-1

There are a number of challenges, including developing new theories for
high-dimensional statistical estimation as well as new formulations
and
computational procedures.
Related problems have received a lot of attention in various research
fields, including
applied math, signal processing, machine learning,\vadjust{\goodbreak}
statistics and optimization.
Rapid advances have been made in recent years.
In view of the growing research activities and their practical importance,
we have organized this special issue of {\sl Statistical Science} with
the goal
of providing
overviews of several topics in modern sparsity analysis and associated
regularization methods.
Our hope is that general readers
will get a broad idea of the field as well as current research directions.

\section{Sparse Modeling and Regularization}

One of the central problem in statistics is linear~re\-gression, where we
consider an
$n \times p$ design matrix~$X$ and an $n$-dimensional response vector
$Y \!\in\!\R^n$ so~that\looseness=-1
%
\begin{equation}\label{eqreg}
Y = X \bbeta+ \varepsilon,
\end{equation}\looseness=0
where $\bbeta\in\R^p$ is the true regression coefficient vector and
$\varepsilon\in\R^n$ is a noise vector.
In the case of $n < p$, this problem is ill-posed because the number of
parameters is more than the number of observations.
This ill-posedness can be resolved by imposing\vspace*{1pt} a
sparsity constraint: that is, by assuming that $\|\bbeta\|_0 \leq s$
for some $s$, where the $\ell_0$-norm of $\bbeta$ is defined as $\|
\bbeta\|_0=|{\supp}(\bbeta)|$, and the support set
of $\bbeta$ is defined as $ \supp(\bbeta):= \{j\dvtx\bbeta_j \neq0\}$.
If $s \ll n$, then the effective number of parameters in (\ref
{eqreg}) is smaller than the number of observations.

The sparsity assumption may be viewed as the classical model selection
problem, where models are indexed
by the set of nonzero coefficients.
The classical model selection criteria such as AIC, BIC or Cp
\mbox{\cite{Akaike73,Mallows73,Schwarz78}} naturally lead
to the so-called $\ell_0$ regularization estimator:
%
\begin{equation}\label{eqreg-L0}
\hbeta^{(\ell_0)} = \arg\min_{\beta\in\R^p} \biggl[ \frac{1}{n} \|X
\beta- Y\|_2^2 + \lambda\|\beta\|_0 \biggr].
\end{equation}
The main difference of modern $\ell_0$ analysis in high-dimensional
statistics and the classical model selection methods is that the
choice of $\lambda$ will be different, and the modern analysis
requires choosing a larger $\lambda$ than that considered
in the classical model selection setting because it is necessary to
compensate for the effect of considering many models in the
high-dimensional setting. The\vadjust{\goodbreak} analysis for $\ell_0$ regularization in
the high-dimensional setting (e.g.,
\cite{ZhZh12-nonconvex} in this issue) employs different techniques
and the results obtained are also different from the classical literature.

The $\ell_0$ regularization formulation leads to a nonconvex
optimization problem that is difficult to solve computationally.
On the other hand, an important requirement for modern high-dimensional problems
is to design computationally efficient and statistically effective algorithms.
Therefore, the main focus of the existing literature is on convex
relaxation methods that use $\ell_1$-regularization (Lasso) to
replace sparsity constraints:
%
\begin{equation}\label{eqreg-L1}
\hbeta^{(\ell_1)} = \arg\min_{\beta\in\R^p} \biggl[ \frac{1}{n} \|X
\beta- Y\|_2^2 + \lambda\|\beta\|_1 \biggr].
\end{equation}
This method is referred to as Lasso \cite{Tibshirani96} in the
literature and its theoretical properties have been intensively studied.
Since the formulation is regarded as an approximation of (\ref
{eqreg-L0}), a key question is how good this approximation is,
and how good is the estimator $\hbeta^{(\ell_1)}$ for estimating
$\bbeta$.

Many extensions of Lasso have appeared in the literature for more
complex problems.
One example is group Lasso \cite{Yuan06JRSS} that assumes that
variables are selected in groups.
Another extension is the estimation of graphical models, where one can
employ Lasso to estimate unknown
graphical model structures \cite{Banerjee-2008,Meinshausen-NSLasso-2006}.
A~third example is matrix regularization, where the concept of sparsity
can be replaced by the concept of low-rankness, and
sparsity constraints become low-rank constraints. Of special interest
is the so-called matrix completion problem,
where we want to recover a matrix from a few observations of the matrix
entries. This problem is encountered in
recommender system applications (e.g., a person buys a book at
\href{http://www.amazon.com}{amazon.com} will be recommended other books purchased by other users
with similar interests), and low-rank matrix factorization is one of
the main techniques for this problem.
Similar to sparsity regularization, using low-rank regularization leads
to nonconvex formulations and, thus, it is natural to
consider its convex relaxation which is referred to as trace-norm (or
Nuclear norm) regularization. The theoretical properties and numerical
algorithms for trace-norm regularization methods have received attention.

\section{Articles in this Issue}

The eight articles in this issue present general over\-views of the state
of the art in a number of different\vadjust{\goodbreak}
topics concerning sparsity analysis and regularization methods.
Moreover, many articles go beyond the current state of the art in
various ways. Therefore, these articles not only give
some high level ideas about the current topics,
but will also be valuable for experts working in the field.
\begin{itemize}
\item
Bach, Jenatton, Mairal and Guillaume (Structured sparsity through
convex optimization, \cite{BJMO12-struct})
study convex relaxations based on structured norms
incorporating further structural prior knowledge.
An extension of the standard $\ell_0$ sparsity model that has received
a lot of attention
in recent years is \textit{structured sparsity}. The basic idea is that
not all sparsity patterns
for $\supp(\bbeta)$ are equally likely. A~simple example is
group sparsity where nonzero coefficients occur together in predefined groups.
More complex structured sparsity models have been investigated
in recent years.
Although the paper by Bach et al. focuses on the convex optimization approach,
they also give an extensive survey of recent developments, including
the use of sub-modular set functions.

\item
van de Geer and M\"uller (Quasi-likelihood and/or robust estimation
in high dimensions, \cite{GeeMul12-quasi})
extend $\ell_1$ regularization methods
to generalized linear models. This involves consideration of loss
functions beyond
the usual least-squares loss and, in particular, loss functions arising
via quasi-likeli\-hoods.

\item
Huang, Breheny and Ma (A selective review of group selection in
high-dimensional regression, \cite{HBM12-group})
provide a detailed review of
the most important special case of structured sparsity, namely, group sparsity.
Their review covers both convex relaxation (or group Lasso)
and approaches based on nonconvex group penalties.

\item
Huet, Giraud and Verzelen (High-dimensional regression with unknown
variance, \cite{GiHuVe12-var})
address issues in high-dimensional regression estimation
connected with lack of knowledge of the error variance.
In the standard Lasso formulation (\ref{eqreg-L1}), the
regularization parameter $\lambda$ is considered as
a tuning parameter that needs to be chosen proportionally to the
standard deviation $\sigma$ of the noise vector.
A natural question is whether it is possible to
automatically estimate $\sigma$ instead of leaving $\lambda$ as a
tuning parameter. This problem has received much attention
and a number of developments have been made in recent years. This paper
reviews and compares several approaches to this problem.\looseness=-1


\item
Lafferty, Liu and Wasserman (Sparse nonparametric graphical models,
\cite{LaLiWa12-graph})\vadjust{\goodbreak}
discuss another important topic in sparsity
analysis, the graphical model estimation problem. While much of the
current work assumes that the data
come from a multivariate Gaussian distribution, this paper goes beyond
the \mbox{standard} practice.
The authors outline a number of possible approaches and introduce more
flexible models for the problem.
The authors also describe some of their recent work, and describe
future research directions.


\item
Negahban, Ravikumar, Wainwright and Yu (A unified framework for
high-dimensional analysis of M-estimators with decomposable
regularizers, \cite{NRWY12-unif})
provide a unified treatment of existing approaches to sparse
regularization. The paper
extends the standard sparse recovery analysis of $\ell_1$ regularized
least squares regression problems
by introducing a general concept of restricted strong convexity. This
allows the authors to study
more general formulations with different convex loss functions and a
class of ``decomposable'' regularization conditions.


\item
Rigollet and Tsybakov (Sparse estimation by exponential weighting,
\cite{RigTsy12-exp})
present a thorough analysis of oracle inequalities
in the context of model averaging procedures, a class of methods which
has its original in the Bayesian literature.
Model averaging is in general more stable than model selection. For
example, in the scenario that
two models are very similar and only one is correct, model selection
forces us to choose one of the models even
if we are not certain which model is true.
On the other hand, a model averaging procedure does not force us to
choose one of the two models, but only to
take the average of the two models.
This is beneficial when several of the models are similar and we cannot
tell which is the correct one. The modern analysis
of model averaging procedures leads to oracle inequalities that are
sharper than the corresponding oracle inequalities for model selection
methods such as Lasso. The authors give an extensive discussion of such
oracle inequalities using an exponentially
weighted model averaging procedure. Such procedures have advantages
over model selection when the underlying models are correlated
and when the model class is misspecified.


\item
Zhang and Zhang (A general theory of concave regularization for high-dimensional
sparse estimation problems, \cite{ZhZh12-nonconvex}) focus on
nonconvex penalties and study a variety of issues related to such penalties.
Although the natural formulation of a sparsity constraint is $\ell_0$
regularization, due to its computational difficulty,
most of the recent literature focuses on the simpler $\ell_1$
regularization method (Lasso) that approximates $\ell_0$ regularization.
However, it is also known that $\ell_1$ regularization is not a very
good approximation to $\ell_0$ regularization.
This leads to the study of nonconvex penalties. The nonconvex
formulations are both harder to analyze statistically and harder to
handle computationally. Some fundamental understanding of
high-dimensional nonconvex procedures
has only started to emerge recently.
Nevertheless, 
some
basic questions have remained unanswered: for example,
properties of the global solution of nonconvex formulations and
whether it is possible to compute the global optimal solution
efficiently under suitable conditions. The authors go a considerable
distance toward providing
a general theory that answers some of these fundamental 
questions.
\end{itemize}




\end{document}